\documentstyle[aps,prl,multicol,epsf]{revtex}
\begin{document}
\draft
%----------------------------------------------------------------
\renewcommand{\narrowtext}{\begin{multicols}{2} \global\columnwidth20.5pc}
\renewcommand{\widetext}{\end{multicols} \global\columnwidth42.5pc}
\multicolsep = 8pt plus 4pt minus 3pt
%----------------------------------------------------------------
\title{Nonlocal Effects on the Magnetic Penetration Depth in {\em
      d}-wave Superconductors}

\author{Ioan Kosztin and Anthony J.\ Leggett}
\address{Department of Physics, University of Illinois at Urbana-Champaign,\\
1110 West Green Street, Urbana, Illinois 61801}

\date{February 20, 1997}

\maketitle
%----------------------------------------------------------------
\begin{abstract}
  We show that, under certain conditions, the low temperature behavior of
  the magnetic penetration depth $\lambda(T)$ of a pure $d$-wave
  superconductor is determined by nonlocal electrodynamics and, contrary to
  the general belief, the deviation $\Delta\lambda(T) =
  \lambda(T)-\lambda(0)$ is proportional to $T^2$ and not $T$.  We predict
  that the $\Delta\lambda(T)\propto T^2$ dependence, due to nonlocality,
  should be observable experimentally in nominally clean high-$T_c$
  superconductors below a crossover temperature $T^* =
  \left(\xi_o/\lambda_o\right)\Delta_o \sim 1$K.  Possible complications due
  to impurities, surface quality and crystal axes orientation are discussed.
\end{abstract}
\pacs{PACS numbers: 
   74.25.Nf, % Response to electromagnetic fields (nuclear magnetic resonance,
             % surface impedance, etc.)
   74.20.Fg, % BCS theory and its development
   74.72.Bk  % Y-based cuprates   
   \hfill {\tt cond-mat/9702199}}

\narrowtext
%--------------------------------------------------------------------------
There is a significant amount of experimental evidence that the pairing state
in the cuprate high temperature superconductors (HTSC) is unconventional,
most probably of $d_{x^2-y^2}$ symmetry
\cite{scalapino95-329,dvh95-515,annett96-375}.  One generic feature of a
layered HTSC with any unconventional order parameter (OP) compatible with the
underlying crystal symmetry is that the OP exhibits line nodes (point nodes
in 2D) on the Fermi surface (FS) and, therefore, gapless quasiparticle
excitations in the corresponding energy spectrum. These low-lying
excitations dominate the low temperature thermodynamics and transport
properties of these materials, and it is expected that the temperature
dependence of the different thermodynamic quantities and transport
coefficients will follow a power law rather than the conventional
exponential behavior\cite{sigrist91-239}.  Direct experimental evidence for
the existence of zeros of the gap function on the FS in HTSC has been found
by angle resolved photo emission spectroscopy (ARPES)
\cite{ding94-1333,ding95-2784}.

In particular, the low temperature behavior of the Meissner penetration
depth $\lambda(T)$ is frequently regarded as an important probe of the
morphology of the magnitude of the anisotropic OP,
$\Delta\left(\hat{\bbox{p}}\right)$, in the cuprates. In conventional
$s$-wave superconductors, the deviation $\Delta\lambda(T)$ of $\lambda(T)$
from its zero temperature value $\lambda(0)$ exhibits activated behavior,
i.e., $\Delta\lambda(T)\propto \exp(-\Delta/T)$ (throughout this paper we
use units in which $k_B=\hbar=1$), reflecting the existence of the isotropic
BCS energy gap $\Delta$ at the FS. In contrast, in a pure $d$-wave
superconductor, or any other unconventional superconductor with nodes in the
gap, the London (local) penetration depth varies linearly with the
temperature, i.e., $\Delta\lambda(T)\propto T$.
Recently, by employing different high precision measurement techniques, such
a linear $T$-dependence of the in-plane $\Delta\lambda_{ab}(T)$ penetration
depth (here the subscript refers to the axes along which the screening
currents flow) has been observed experimentally in the Meissner state of
several HTSC systems, such as: high quality single crystals of
YBa$_2$Cu$_3$O$_{7-\delta}$ (YBCO)
\cite{hardy93-3999,zhang94-2484,mao95-3316} and Bi$_2$Sr$_2$CaCu$_2$O$_8$
(BSCCO) \cite{jacobs95-4516,lee96-735,waldmann96-11825}, magnetically
aligned powders of crystalline HgBa$_2$Ca$_2$Cu$_3$O$_{8+\delta}$
\cite{panagopoulos96-r2999} and high quality YBCO thin films
\cite{froehlich94-13894,froehlich96-467,fuchs96-14745}.
However, below a certain sample dependent temperature $T_{\text{imp}}^*$ the
linear $T$-dependence of the penetration depth in HTSC crosses over into a
higher power law, most probably $T^2$. 
In the $d$-wave scenario of HTSC, the origin of the $\lambda(T)\propto T^2$
dependence has been explained by the presence of non-magnetic impurities
which scatter in the unitary limit \cite{hirschfeld93-4219}.  In this strong
scattering limit a small amount of impurities can induce a finite residual
density of states at the Fermi level which is sufficient to change the
temperature dependence of the penetration depth from $T$ to $T^2$ without
lowering significantly the transition temperature. A direct experimental
confirmation of such a crossover between pure and impurity dominated regimes
was reported by Bonn {\em et al.} \cite{bonn94-4051}.

The purpose of this Letter is to show that at very low temperatures
nonlocality may play an important role in the electromagnetic response of a
$d$-wave superconductor (or any other unconventional superconductor with
nodes in the gap), leading to a $\Delta\lambda(T)\propto T^2$ dependence even
in the clean limit. Thus, besides impurities, nonlocality represents a
second mechanism which leads to a $T^2$ dependence of the penetration depth
sufficiently close to $T=0\,$K.
To the best of our knowledge, all theoretical calculations and
interpretations of the experimental measurements of the penetration depth
in HTSC performed so far assume the validity of local electrodynamics
(London limit)\cite{sauls}. At first sight this is reasonable, since the 
zero temperature London penetration depth $\lambda_o= \sqrt{mc^2/4\pi ne^2}$
is much larger than the corresponding coherence length $\xi_o$ in these
materials.
(In contrast to the penetration depth, which can be measured more or less
directly, the coherence length $\xi_o$ cannot be determined experimentally
and, in fact, is estimated in terms of the maximum value of the
anisotropic gap function $\Delta_o = \text{max} \{\Delta(\hat{\bbox{p}})\}$
by using the usual BCS expression $\xi_o = v_F/\pi\Delta_o$.)
However, in the case of a clean, anisotropic superconductor it is more
appropriate to introduce an anisotropic coherence length
$\xi\left(\hat{\bbox{p}}\right) \equiv v_F/
\left|\Delta\left(\hat{\bbox{p}}\right)\right|$. If the anisotropic OP has
nodes on the FS, it is clear that sufficiently close to the nodes
$\xi\left(\hat{\bbox{p}}\right)/\lambda_o = \left(\xi_o/\lambda_o\right)
\Delta_o/\left|\Delta\left(\hat{\bbox{p}}\right)\right| \gtrsim 1$ holds
and, therefore, the contribution of these regions of the FS to the
penetration depth $\lambda(T)$ must be determined by using nonlocal
electrodynamics.  The large value of $\lambda_o/\xi_o$ guarantees that the
applicability of local electrodynamics is violated only on a very small
fraction, of order $\alpha_o\equiv\xi_o/\lambda_o$, of the FS.  Since the
whole FS contributes to the zero temperature penetration depth $\lambda(0)$,
one expects no significant corrections to this quantity due to nonlocal
effects.  On the other hand, the low temperature dependence of $\lambda(T)$
must be dominated by nonlocal effects because this dependence is determined
by a small region of the FS which is concentrated around the nodes of the
OP.
The crossover temperature below which nonlocal effects are important is
given by $T^*=\alpha_o\Delta_o$. Indeed, since the range of $\hat{\bbox{p}}$
values corresponding to the thermally excited quasiparticles at a given
temperature $T$ is determined by the condition
$\left|\Delta\left(\hat{\bbox{p}}\right)\right| \lesssim T$, for $T<T^*$ one
obtains $\xi\left(\hat{\bbox{p}}\right)/\lambda_o =
\alpha_o\Delta_o/\left|\Delta\left(\hat{\bbox{p}}\right)\right| \gtrsim
T^*/T > 1$. For $T\gg T^*$ the local limit is applicable. As a typical
example consider a YBCO single crystal with $\Delta_o\approx 250$K,
$\xi_o\approx 14$\AA\ and $\lambda_o\approx 1400$\AA; this yields
$\alpha_o\approx 10^{-2}$ and $T^*\approx 2.5$K.

To demonstrate the effect of nonlocal electrodynamics on $\lambda(T)$ let us
consider the case when a weak, uniform and static magnetic field
$\bbox{H}=\nabla\times\bbox{A}$ is applied along the $c$-axis of a
semi-infinite HTSC with a plane surface which is perpendicular to the
$b$-axis. 
For this particular geometry, both the vector potential $\bbox{A}$
and the screening supercurrent density $\bbox{j}$ are oriented parallel to
the $a$-axis, while the direction of penetration is along the $b$-axis. 
We model the HTSC as a quasi two-dimensional $d$-wave superconductor
in which the motion of the electrons is confined for our purposes to the
Cu-O planes. 
In principle, to calculate $\lambda(T)$, one must first solve
self-consistently the relevant Maxwell equation
$\nabla\times\nabla\times\bbox{A} = \left(4\pi/c\right)\bbox{j}$ together
with the equation which relates $\bbox{j}$ to $\bbox{A}$, subject to some
properly chosen boundary conditions. In a weak magnetic field (Meissner
state) linear response theory yields for our geometry $j(y) = - \int\!
dy'\,K\left(y,y'\right)\, A\left(y'\right)$, where $y$ is the coordinate
along the $b$-axis ($y=0$ gives the position of the boundary) and the
nonlocal electromagnetic response kernel $K\left(y,y'\right)$ must be
calculated by using some microscopic theory.
Once $H(y)$ is determined, the penetration depth can be calculated according
to the standard definition (valid for a semi-infinite superconductor with
plane boundary) $\lambda = H(0)^{-1} \int_0^{\infty}\! dy\,H(y)$.
Furthermore, we assume that the boundary reflects the electrons either
specularly or diffusively. In both these limiting cases $\lambda(T)$ can be
expressed in terms of the Fourier transform of the bulk response kernel
$K(q;T)$. For a specular boundary one has \cite{tinkham}

\begin{equation}
  \label{eq:1}
  \frac{\lambda_{\text{spec}}(T)}{\lambda_o} \;=\; \frac{2}{\pi}
  \int_0^{\infty}\!  \frac{d\tilde{q}}{\tilde{q}^2 +
    \tilde{K}(\tilde{q};T)}\;, 
\end{equation}
while for a diffuse boundary \cite{tinkham}
\begin{equation}
  \label{eq:2}
  \frac{\lambda_{\text{diff}}(T)}{\lambda_o} \;=\;
  \pi\left\{\int_0^{\infty}\!d\tilde{q}
      \ln\left[1+\tilde{K}\left(\tilde{q};T\right)/\tilde{q}^2\right]
    \right\}^{-1}\;,
\end{equation}
where the dimensionless quantities $\tilde{q}$ and $\tilde{K}$ are given by
$\tilde{q}=q\lambda_o$ and $\tilde{K}=\left(4\pi\lambda_o^2/c\right) K$,
respectively. 

For a weak-coupling, anisotropic superconductor the nonlocal bulk response
kernel is similar to the corresponding expression for a conventional
$s$-wave superconductor \cite{agd} and can be written as
\begin{equation}
  \label{eq:3}
  \tilde{K}\left(\tilde{q};T\right) \;=\; 2\pi T \sum_{n=-\infty}^{\infty}
  \left\langle \hat{p}_{||}^2 \frac{\Delta_p^2}{\sqrt{\omega_n^2+\Delta_p^2}
    \left(\omega_n^2+\Delta_p^2+\alpha^2\right)} \right\rangle\;,
\end{equation}
where $\omega_n$ are fermionic Matsubara frequencies, $\Delta_p\equiv
\Delta\left(\hat{\bbox{p}}\right)$, $\hat{p}_{||}$ is the projection of
$\hat{\bbox{p}}$ on the boundary, $\langle\ldots\rangle$ means averaging
over the circular 2D Fermi surface, and $\alpha = \left(qv_F/2\right)
\hat{\bbox{q}}\hat{\bbox{p}}$. Here $\hat{\bbox{q}}$ is a unit vector
perpendicular to the boundary and it gives the direction in which the
penetration of the magnetic field takes place. Note that in a different
geometry where the boundary is parallel to the $a$-$b$ plane ($\bbox{H}$
parallel to the boundary), the direction of penetration $\hat{\bbox{q}}$
would be along the $c$-axis, i.e., perpendicular to $\hat{\bbox{p}}$,
yielding $\alpha = 0$. Thus, we may conclude that the effect of nonlocal
electrodynamics on $\lambda_{ab}(T)$ is relevant only when $\bbox{H}$ is
parallel to the $c$-axis.
Furthermore, at sufficiently low temperatures, the OP in Eq.~(\ref{eq:3})
can be approximated with its limiting expression close to the nodes, i.e.,
$\Delta_p = \Delta_o\Phi\left(\hat{\bbox{p}}\right) \approx \Delta_o
\Phi'(0)\,\varphi$, where $\varphi$ is the angular deviation of
$\hat{\bbox{p}}$ from the given node direction in the basal plane. In the
case of a model $d$-wave OP with $\Phi\left(\hat{\bbox{p}}\right) =
\hat{p}_x^2-\hat{p}_y^2$ one has $\Phi'(0)=2$. 

Let us calculate first the nonlocal correction to the zero temperature
penetration depth $\lambda(0)$. For $T=0$ the frequency sum in
(\ref{eq:3}) goes into an integral which can be evaluated exactly with the
result 
\begin{equation}
   \label{eq:4}
   \tilde{K}\left(\tilde{q};0\right) \;=\; 1 - \left\langle 2
     \hat{p}_{||}^2 \left[1 - \frac{\sinh^{-1}
         \left(\alpha/\Delta_{\bbox{p}}\right)}{
         \left(\alpha/\Delta_{\bbox{p}}\right)
         \sqrt{1+\left(\alpha/\Delta_{\bbox{p}}\right)^2}}
     \right]\right\rangle\;.
\end{equation}
The average over the FS in (\ref{eq:4}) can be evaluated analytically for
both London (local) and Pippard (extreme nonlocal) limits. In general, for
arbitrary $\tilde{q}$ values, $\tilde{K}\left(\tilde{q};0\right)$ must be
calculated numerically.  In the London limit, when
$\alpha_o\tilde{q}=q\xi_o\ll 1$, one obtains
$\tilde{K}\left(\tilde{q};0\right) = 1 - \left(\pi^2\sqrt{2}/16\right)
\alpha_o\tilde{q}$, while in the Pippard limit, when $\alpha_o\tilde{q}\gg
1$, one has $\tilde{K}\left(\tilde{q};0\right) = (2/3)
\ln\left(\alpha_o\tilde{q}\right)/\left(\alpha_o\tilde{q}\right)^2$.
Note that in both limiting cases the response kernel for a $d$-wave
superconductor decreases with $\tilde{q}$ more rapidly than for a
conventional $s$-wave superconductor\cite{tinkham}.
Now the correction to the zero temperature penetration depth due to
nonlocality can be obtained from Eqs.~(\ref{eq:1}-\ref{eq:2}) for
both specular and diffuse boundaries. The results are:
$\Delta\lambda_{\text{spec}}(0)/\lambda_o =
\lambda_{\text{spec}}(0)/\lambda_o - 1 = \pi\sqrt{2}\,\alpha_0/16$, and
$\Delta\lambda_{\text{diff}}(0)/\Delta\lambda_{\text{spec}}(0) =
\ln\left(\alpha_o^{-2}\right)/2 \approx 4.6$.  Thus for both type of
boundaries, due to the very small value of $\alpha_o$, the nonlocal
correction to $\lambda(0)$ is less than 1\% and therefore it can be
obviously neglected, especially because this correction is situated within
the experimental errors of the most accurate measurements of the absolute
value of the penetration depth.

We turn now to calculate $\Delta\lambda(T)$. At low temperatures,
$\delta\tilde{K}\left(\tilde{q};T\right) \equiv
\tilde{K}\left(\tilde{q};T\right) - \tilde{K}\left(\tilde{q};0\right)$
represents a small correction to the zero temperature response kernel
$\tilde{K}\left(\tilde{q};0\right)$.  Therefore, by using our previous
result $\lambda(0)\approx\lambda_o$, from Eqs.~(\ref{eq:1}-\ref{eq:2}) one
obtains

\begin{equation}
  \label{eq:1a}
  \frac{\Delta\lambda_{\text{spec}}(T)}{\lambda_o} \;=\; \frac{2}{\pi}
  \int_0^{\infty}\!\!d\tilde{q}\,
  \frac{-\delta\tilde{K}\left(\tilde{q};T\right)}{\left(\tilde{q}^2 +
      1\right)^2}\;,
\end{equation}
and
\begin{equation}
  \label{eq:2a}
  \frac{\Delta\lambda_{\text{diff}}(T)}{\lambda_o} \;=\; \frac{1}{\pi}
  \int_0^{\infty}\!\!d\tilde{q}\,
  \frac{-\delta\tilde{K}\left(\tilde{q};T\right)}{\tilde{q}^2+1}\;.
\end{equation}

Furthermore, a convenient expression for $\delta\tilde{K}$ can be obtained
by evaluating the Matsubara sum in Eq.~(\ref{eq:3}) by means of complex
contour integration

\[
  -\delta\tilde{K}\left(\tilde{q};T\right) \;=\; 2 \int_0^{\infty}\!
  f(\omega)\, d\omega\, 
\]
\begin{equation}
  \label{eq:5}
  \times\left\langle 2\hat{p}_{||}^2\, \text{Re}
  \frac{\Delta_p^2}{\sqrt{\omega^2-\Delta_p^2}
    \left(\Delta_p^2+\alpha^2-\omega^2\right)} \right\rangle\;.
\end{equation}

\noindent Note that in the $\alpha\rightarrow 0$ limit one recovers the
familiar local limit expression for $\delta\tilde{K}$
\cite{hirschfeld93-4219}.
Due to the presence of the Fermi function $f(\omega)$ in (\ref{eq:5}) the
main contribution to the frequency integral comes from the interval
$\omega\lesssim T$. Therefore, in the average over the FS the relevant
regions are determined by $|\Delta_p|\le\omega\lesssim T$ and are obviously
located around the nodes of the OP. By using the expression for the OP close
to a node one arrives, after some straightforward algebra, at the following
result

\begin{equation}
  \label{eq:6}
  \delta\tilde{K}\left(\tilde{q};T\right) \;=\; \delta\tilde{K}(0;T)\,
  F\left(\frac{\tilde{q}}{t}\right)\;, 
\end{equation}
where $t\equiv T/T^*$, $\delta\tilde{K}(0;T) = -2\ln2\,T/\Delta_o$ is the
well known local limit expression of $\delta\tilde{K}$ for a $d$-wave
superconductor \cite{scalapino95-329}, the expression

\begin{equation}
  \label{eq:7}
  F(z) \;=\; 1-\frac{1}{\ln2} \int_0^{\frac{\pi\sqrt{2}}{4}z}\! dx\, f(x)\,
  \sqrt{1-\frac{8}{\pi^2}\,\left(\frac{x}{z}\right)^2}
\end{equation}
is a universal function, and $f(x)=\left(e^x+1\right)^{-1}$. It is
remarkable that, within the above mentioned approximations, the kernel
$\delta\tilde{K}$ depends only on the ratio $\tilde{q}/t$ and not
separately on its two arguments. 
In order to make further analytical progress it is desirable to approximate
$F(z)$ by a simpler expression. A careful quantitative analysis of
Eq.~(\ref{eq:7}) motivates the following reasonable approximation: $F(z)
\approx 1 - c_1 z$, for $z < 2$, and $F(z) \approx c_o/z^2$, for $z > 2$,
where $c_o = 6\zeta(3)/\pi^2\ln2 \approx 1.05$, and $c_1 =
\left(1-c_o/4\right)/2 \approx 0.37$. Note that $c_1$ is somewhat smaller
than the absolute value of the slope of $F(z)$ at the origin, i.e.,
$|F'(0)| = \pi^2\sqrt{2}/32\,\ln2 \approx 0.63$.
The temperature dependence of the penetration depth can now be calculated by 
inserting (\ref{eq:6}) in Eqs.~(\ref{eq:1a}-\ref{eq:2a}).

For $t\gg 1$ (i.e., $T\gg T^*$) and for a specularly reflecting boundary one
obtains $\Delta\lambda_{\text{spec}}(T)/\lambda_o = \ln
2\,\left(T/\Delta_o\right) - \left(\pi\sqrt{2}/16\right)\,\alpha_o + {\cal
  O}(1/t)$.  Here, the leading term is the well known linear in $T$ local
expression for $\Delta\lambda(T)$ for a $d$-wave superconductor, i.e.,
$\Delta\lambda_L(T) = \ln 2\, \left(T/\Delta_o\right)\,\lambda_o$.  The
second, small negative constant term in the expression of
$\Delta\lambda_{\text{spec}}(T)$ is due to nonlocality and shows clearly
that the linear $T$-dependence cannot extend all the way down to $T=0\,$K;
it must cross over to a higher power law at some $T\sim T^*$.
In the case of a diffuse boundary one obtains a similar result, namely 
$\Delta\lambda_{\text{diff}}(T) = \Delta\lambda_L(T) - 
\left(\pi\sqrt{2}/16\right)\,\alpha_o\lambda_o\, \ln t + {\cal O}(1/t)$.  Note 
that the magnitude of the nonlocal correction to the local penetration depth 
is larger than in the case of the specular boundary by a factor of $\ln t$.

\begin{figure}[htbp]
  \centerline{\epsfxsize=3.4in\epsffile{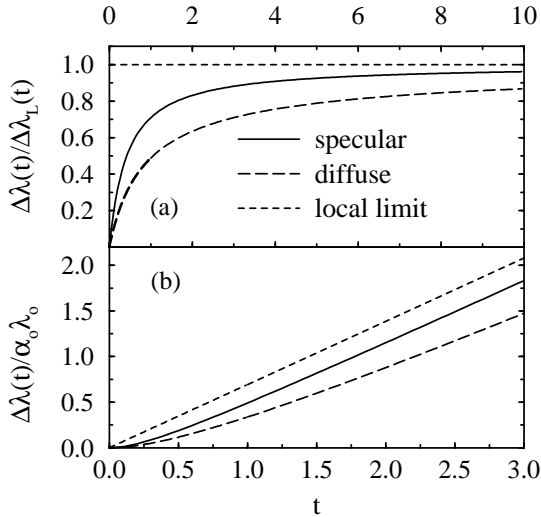}}
  \caption{Plot of $\Delta\lambda(T)$ [in units (a) $\Delta\lambda_L(T)$, and
    (b) $\alpha_o\lambda_o$, respectively] vs.\ $t=T/T^*$ for
    both specular (solid line) and diffuse (long-dashed line) boundary. For
    comparison, the local limit result is also shown (dashed line).}
  \label{fig}
\end{figure}

In the opposite limit $t\ll 1$ (i.e., $T\ll T^*$) one obtains for a specular
boundary $\Delta\lambda_{\text{spec}}(T) = \beta\, \Delta\lambda_L(T)\,
T/T^* \propto T^2$, where $\beta =
8\left(1-c_1+c_o/4\right)/\pi \approx 2.2$.  Thus, due to nonlocal
electrodynamics, for $T\ll T^*$ the temperature dependence of a {\em pure\/}
$d$-wave superconductor is proportional to $T^2$ and not $T$, regardless of
how small is $\alpha_o = \xi_o/\lambda_o$.  This conclusion is one of the
main results of the present paper.  It should be noted that the above value
for the coefficient $\beta$ is just an approximation; a more accurate value
of $\beta$ can be obtained by approximating $F(z)$ by a polynomial of degree
$N>1$ for $z<z_o$, and by its large $z$ asymptotic form for $z>z_o$, where
$z_o$ is a conveniently chosen value.  By reevaluating the integrals in
$\Delta\lambda(T)$ one obtains again the leading term proportional to $T^2$
but with a slightly different numerical value for $\beta$.
A similar calculation in the case of a diffuse boundary yields 
$\Delta\lambda_{\text{diff}}(T) \approx \Delta\lambda_{\text{spec}}(T)/2$.
Note that in both limiting cases the inequalities 
$\Delta\lambda_{\text{diff}}(T) < \Delta\lambda_{\text{spec}}(T) < 
\Delta\lambda_L(T)$ hold and imply that the deviation from the corresponding 
local result is larger for a diffuse boundary than for a specular one (see 
also Fig.~\ref{fig}).

For arbitrary temperatures $\Delta\lambda(T)$ must be calculated numerically 
by employing the exact expression (\ref{eq:7}) for the function $F(z)$.  In 
Fig.~\ref{fig}a the ratio $\Delta\lambda(T)/\Delta\lambda_L(T)$ is plotted, 
for both specular and diffuse boundaries, as a function of the reduced 
temperature $t$.  The deviation from the standard result obtained in the local 
limit is evident.  The clear linear dependence in the vicinity of the origin 
indicates a quadratic $T$-dependence of $\Delta\lambda(T)$.  For $t\gg 1$, 
$\Delta\lambda(T)$ approaches asymptotically its local limit (minus a small 
constant correction of order $\alpha_o$).
Note that the deviation of $\Delta\lambda(T)$ from the corresponding local 
expression is much more pronounced for a diffuse boundary then for a
specular one.
The same $\Delta\lambda(T)$, this time in units of $\alpha_o\lambda_o$, is
shown as a function of $t=T/T^*$ in Fig.~\ref{fig}b. The deviation from
linearity becomes visible around $t=1$ ($t=2$) for the specular (diffuse)
boundary. For the numerical example considered above for a clean YBCO single
crystal ($\alpha_o\lambda_o = \xi_o \approx 14$\AA) one finds that the
deviation from $\Delta\lambda(T) \propto T$ takes place somewhere between 2
to 5 K, depending on the surface quality of the crystal.  Such a crossover
is seen experimentally in nominally clean YBCO crystals \cite{bonn96-7}; in
the existing literature it has been attributed to impurities which scatter
in the unitary limit \cite{hirschfeld93-4219}.

In principle, there is a simple experimental test to determine whether this
crossover in $\Delta\lambda(T)$ is due to nonlocal electrodynamics or to
impurities. The main idea is to estimate experimentally the crossover
temperature in $\Delta\lambda_{ab}(T)$ for the same nominally clean HTSC for
two different magnetic field orientations: (i) $\bbox{H}$ parallel to the
$c$-axis, and (ii) $\bbox{H}$ parallel to the $a$-$b$ plane. As we have already
mentioned, nonlocality is expected to be relevant only when the applied
magnetic field is oriented parallel to the $c$-axis (so that the penetration
direction lies in the $a$-$b$ plane), while the effect of
impurities should not depend on the orientation of the field. Thus, if $T^*$
is noticeably smaller in case (ii) than in case (i) one may conclude that
the observed effect is mainly due to nonlocal electrodynamics and not to
impurities. Otherwise, the conclusion is that nonlocal effects are in fact
completely masked by impurities. 

In conclusion, we have shown that nonlocal electrodynamics dominate the low
temperature behavior of the in-plane magnetic penetration depth of a clean
$d$-wave high-$T_c$ superconductor. At temperatures $T\ll T^*\sim 1$K the
penetration depth $\lambda(T)$ has a quadratic temperature dependence, while
above the crossover temperature $T^*$, but still well below $T_c$,
$\lambda(T)$ has the well known linear $T$-dependence. Thus, nonlocality
represents a second possible mechanism, beside strongly scattering
impurities, which may account for the experimentally observed deviation from
the linear $T$-dependence of the penetration depth at the lowest measured
temperatures in nominally clean HTSC. A simple experiment to probe the
viability of this mechanism has been proposed.

%----------------------------------------------------------------
We are grateful to N.\ Goldenfeld and J.\ A.\ Sauls for helpful discussions.
One of us (A.J.L.) thanks G.\ E.\ Volovik for a conversation in which the
possibility of this effect first emerged.
This work was supported by the National Science Foundation (DMR 91-20000)
through the Science and Technology Center for Superconductivity.
%----------------------------------------------------------------

\widetext
%----------------------------------------------------------------
\end{document}